\documentclass[11pt]{article}

\usepackage[usenames,dvipsnames,svgnames]{xcolor}

\RequirePackage{hyperref}
\hypersetup{colorlinks=true, urlcolor=Blue, citecolor=Green, linkcolor=BrickRed, breaklinks, unicode}

\usepackage[letterpaper,margin=1in]{geometry}
\usepackage{microtype}
\usepackage{tikz}
\usepackage{enumitem}
\usepackage{amsmath,amssymb,amsthm}

\usetikzlibrary{arrows.meta,calc}
\setlist[itemize]{label=$\bullet$,leftmargin=1.5em,itemsep=2pt,topsep=4pt}
\newtheorem{theorem}{Theorem}
\newtheorem{lemma}[theorem]{Lemma}

\newcommand{\dist}{d}
\newcommand{\cut}{\partial}

\definecolor{sideblue}{RGB}{48,97,151}
\definecolor{sideorange}{RGB}{183,102,36}
\definecolor{cutgreen}{RGB}{42,121,91}
\hypersetup{pdftitle={\~{O}ptimal Facial Distance Labeling in Planar Graphs},pdfsubject={A simplified proof of the one-face distance labeling theorem}}
\title{\~{O}ptimal Facial Distance Labeling in Planar Graphs}

\author{%
Noy Rotbart\\
   Churney Copenhagen, Denmark\\
 \texttt{\href{mailto:noy@churney.io}{noy@churney.io}}\\
 \and Oren Weimann\\
 University of Haifa, Israel\\
 \texttt{\href{mailto:oren@cs.haifa.ac.il}{oren@cs.haifa.ac.il}}
}

\hypersetup{pdfauthor={Oren Weimann and Noy Rotbart}}
\date{}
\begin{document}
\begin{center}
{\LARGE \~{O}ptimal Facial Distance Labeling in Planar Graphs\par}
\vspace{.65em}
\makeatletter
{\normalsize\begin{tabular}[t]{c}\@author\end{tabular}\par}
\makeatother
\end{center}
\begin{abstract}
We show how to assign labels of size $\tilde O(n^{1/3})$ to the vertices of a face $F$ in an $n$-vertex unweighted undirected planar graph $G$ s.t one can deduce the distance between any two vertices of $F$ given their labels alone. 
This matches (up to log factors) the $\Omega(n^{1/3})$ lower bound of Gavoille, Peleg, P{\'e}rennes, Raz from 2004. Prior to our work, the smallest labels were of size $O(n^{1/2})$.
\end{abstract}

\section{Introduction}
A distance labeling scheme assigns a short label to each vertex so that the distance between any two  vertices can be deduced from their labels alone.
For $n$-vertex unweighted undirected planar graphs, Gavoille, Peleg, P\'erennes and Raz~\cite{gppr} proved an $\Omega(n^{1/3})$ lower bound and an $\tilde O(n^{1/2})$ upper bound on the label size.\footnote{The upper bound was improved to $O(n^{1/2})$ (i.e., without log factors) by Gawrychowski and Uzna{\'n}ski~\cite{gu}.}. Closing this gap is one of the most interesting open problems in distance labeling.

It turns out that the lower-bound construction of Gavoille et al.~\cite{gppr} places all vertices of interest on a single face $F$.
That is, it is actually an $\Omega(n^{1/3})$ lower bound for the case where we only need to assign labels to the vertices of a single face $F$.  
We show that this case actually admits distance labels of $\tilde O(n^{1/3})$ bits, resolving this case up to logarithmic factors.

\begin{theorem}\label{thm:main}
Let $G$ be an $n$-vertex unweighted undirected planar graph with a fixed embedding, and let $F$ be a face. The vertices of $F$ admit  distance labels of size $O(n^{1/3}\log^2 n)$.
\end{theorem}

At the heart of the proof is the well-known fact that distances between
vertices of the same face form a {\em unit-Monge}
matrix. After a standard reduction, the difference between two consecutive
rows changes sign at most once (at what we call a 
{\em switch point}). Together with the first row and column, these points
determine the matrix: retrieving an entry reduces to a dominance-counting
query on the switch points. The labeling scheme of Abboud, Gawrychowski,
Mozes and Weimann~\cite{agmw} uses this to obtain labels of size $\widetilde O(\sqrt{t})$
 for a $t\times t$ unit-Monge matrix, which suffices when
$t\le n^{2/3}$. However, a face $F$ may contain $\Theta(n)$ vertices,
so these matrices need not be small. Our contribution is a recursive
decomposition of large facial distance matrices. Each switch point
corresponds to two consecutive vertices of $F$, which partition the
graph according to which vertex is closer. The edges between the two
parts form a cut. An averaging argument allows us to choose a switch
point $(i,j)$ whose cut is small and whose horizontal and vertical split balances
the number of switch points. We then partition the matrix into four parts at row $i$ and column $j$. We store distances to the (few) cut-edge endpoints to handle vertex pairs on opposite sides of the cut (i.e., to handle two out of the four parts without recursion), and recurse on the
remaining two blocks. Once a block has few switch points, we apply the
dominance-counting scheme of~\cite{agmw}.

\paragraph{AI disclosure.} 
The proof was entirely discovered by Claude Fable 5.1. The authors clarified and simplified the proof and take full responsibility for its correctness. 

\newpage 
\section{Preliminaries}\label{sec:reduction}
We review two standard lemmas  (see e.g. \cite{Oren_patterns,mww}). 

\begin{lemma}\label{lem:reduction}
It suffices to prove the upper bound for connected bipartite planar graphs whose designated face $F$ is bounded by a simple cycle.
\end{lemma}
\begin{proof}
For a disconnected graph, prefix each label with its component identifier; vertices in different components have distance $\infty$. Within each component, the vertices of $F$ lie on a common face. Isolated vertices are immediate, so assume that $G$ is connected and has an edge.

Subdivide every edge once. This preserves the faces, doubles distances between original vertices, and makes the graph bipartite. Its size is $O(n)$ by planarity. Let $(w_0,\ldots,w_{L-1})$ be the boundary walk of the designated face; vertices may repeat.

Draw a ring inside this face: for each boundary occurrence $w_i$, add a new vertex $z_i$ and an edge $z_iw_i$, and join $z_i$ to $z_{i+1}$ by a new path of length three, with indices modulo $L$. These paths form a simple facial cycle. The graph stays bipartite: color $z_i$ opposite to $w_i$, and color each new path alternately. The ring adds $3L=O(n)$ vertices.

No new path improves distances through the old graph. Indeed, a ring segment consisting of $t\ge1$ length-three paths has length $3t$; replacing it by its two spokes and the corresponding boundary walk has length $t+2\le3t$. Thus, for distinct ring vertices,
\[
  \dist_{\rm new}(z_i,z_j)=\dist_{\rm subdivided}(w_i,w_j)+2.
\]
Choose one occurrence for each original face vertex and give it the label of the corresponding $z_i$. Include an original vertex identifier to handle equal-vertex queries. For distinct vertices, return $(\delta-2)/2$, where $\delta$ is the decoded ring distance.
\end{proof}

From now on, $G$ is connected and bipartite, $C$ is a simple cycle bounding a face, and $m=|E(G)|=O(n)$. For every edge $uv$ and vertex $x$,
\begin{equation}\label{eq:parity}
  |\dist(x,u)-\dist(x,v)|=1:
\end{equation}
the difference is at most one by the triangle inequality and is odd by bipartiteness.

Consider the vertices of $C$ in cyclic order. 

\begin{lemma}\label{lem:consecutive}
For every edge $uv$, the vertices of $C$ closer to $u$ than to $v$ are consecutive on $C$.
\end{lemma}
\begin{proof}
Let $U$ and $V$ be the sets of all graph vertices closer to $u$ and to $v$, respectively. Equation~\eqref{eq:parity} makes them a partition. Every shortest path from $x\in U$ to $u$ stays in $U$: for a vertex $z$ on that path,
\[
 \dist(v,z)\ge \dist(v,x)-\dist(z,x)
 =\dist(u,x)+1-\dist(z,x)=\dist(u,z)+1.
\]
Similarly, every shortest path from a vertex in $V$ to $v$ stays in $V$.

If $U\cap V(C)$ were not consecutive on $C$, four vertices $x_1,x_2,x_3,x_4$ would occur in this cyclic order, with $x_1,x_3\in U$ and $x_2,x_4\in V$. Paths through $u$ give an $x_1$--$x_3$ path in $U$; paths through $v$ give an $x_2$--$x_4$ path in $V$. These paths are disjoint. This is impossible for alternating vertices of a facial cycle: viewing its face as the exterior, the Jordan curve theorem forces the paths to meet.
\end{proof}

\section{Bisectors}\label{sec:bisectors}

Fix two vertex-disjoint nonempty subpaths of $C$,
\[
 A=(a_0,\ldots,a_{p-1}),\qquad B=(b_0,\ldots,b_{p'-1}).
\]
 The vertices are listed in the same cyclic direction, with $A$ followed by $B$ (possibly with gaps). Define
\[
 M(i,j)=\dist(a_i,b_j),\quad
 X_i=\{x:\dist(x,a_{i+1})<\dist(x,a_i)\},\quad
 \pi(i)=|X_i\cap V(B)|.
\]
 For a set of vertices $X$, we denote the {\em bisector} $\cut(X)$ as the edges of $G$ with exactly one endpoint in $X$.

\begin{lemma}\label{lem:cuts}
The sets $X_i$ have the following properties.
\begin{enumerate}[label=(\roman*),leftmargin=2em,itemsep=3pt]
\item $X_i\cap V(B)=\{b_0,\ldots,b_{\pi(i)-1}\}$, and
\[
 M(i+1,j)-M(i,j)=
 \begin{cases}-1,&j<\pi(i),\\ +1,&j\ge\pi(i).\end{cases}
\]
\item If $0<\pi(i)<p'$, then $X_i\cap V(A)=\{a_{i+1},\ldots,a_{p-1}\}$.
\item $\displaystyle\sum_{i=0}^{p-2}|\cut(X_i)|\le2m$.
\end{enumerate}
\end{lemma}
\begin{proof}
By Lemma~\ref{lem:consecutive}, $X_i\cap V(C)$ consists of consecutive vertices, contains $a_{i+1}$, and excludes its predecessor $a_i$. It is therefore a prefix of the cyclic list starting at $a_{i+1}$ and ending at $a_i$. This proves (i). If the prefix contains some but not all of $B$, it contains precisely the suffix $a_{i+1},\ldots,a_{p-1}$ of $A$, proving (ii).
For (iii), fix an edge $uv$ and set $g(i)=\dist(u,a_i)-\dist(v,a_i)$. The edge belongs to $\cut(X_i)$ exactly when the two differences
\[
 \dist(u,a_{i+1})-\dist(u,a_i),\qquad
 \dist(v,a_{i+1})-\dist(v,a_i)
\]
are unequal, or equivalently when $g(i+1)\ne g(i)$. By Lemma~\ref{lem:consecutive}, the vertices of $C$ where $g=-1$ are consecutive. Thus $g$ changes at most twice along $A$, and each edge is counted at most twice.
\end{proof}

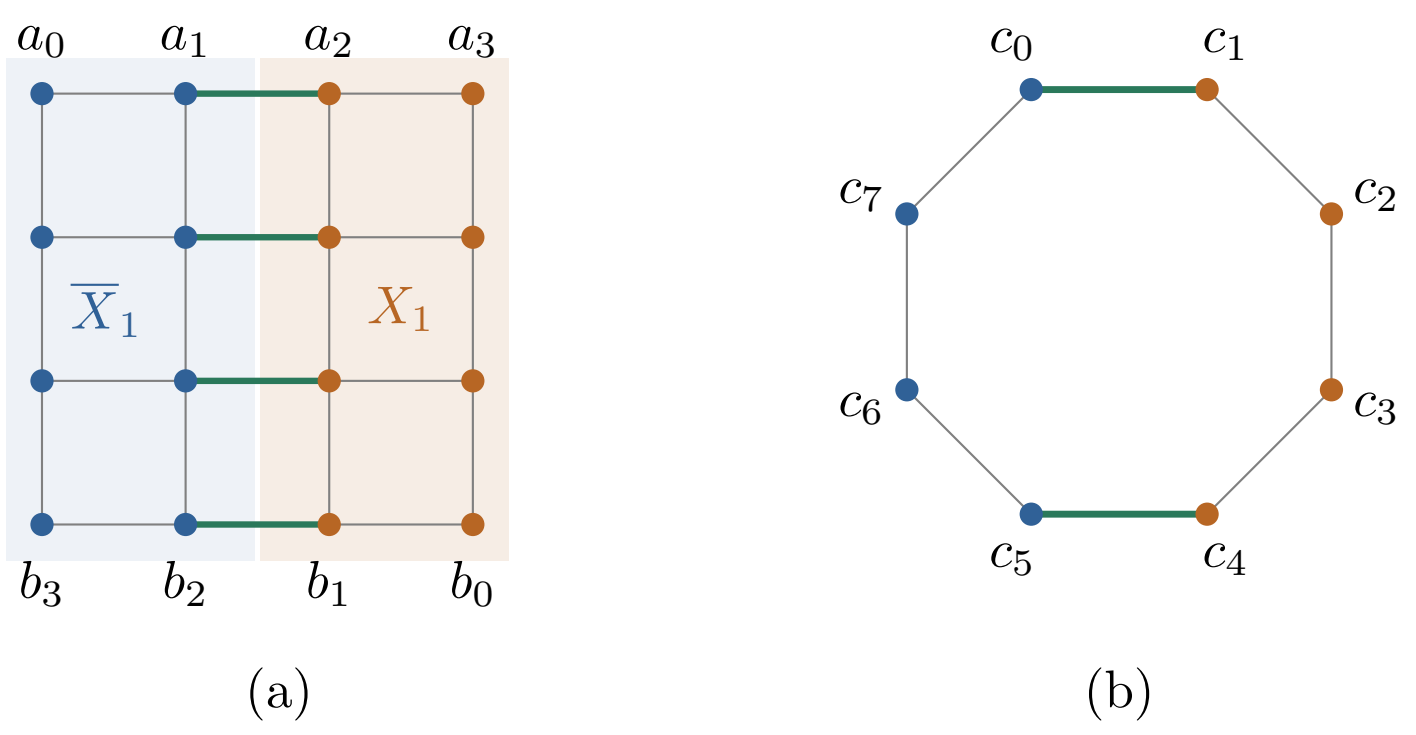
\begin{figure}[t]
\centering
\begin{tikzpicture}[scale=.95,every node/.style={font=\small}]
  \fill[sideblue!8](-.25,-.25) rectangle (1.48,3.25);
  \fill[sideorange!12](1.52,-.25) rectangle (3.25,3.25);
  \foreach \x in {0,1,2,3}{\draw[gray] (\x,0)--(\x,3);}
  \foreach \y in {0,1,2,3}{\draw[gray] (0,\y)--(3,\y);\draw[cutgreen,line width=1.2pt](1,\y)--(2,\y);}
  \foreach \x in {0,1}{\foreach \y in {0,1,2,3}{\fill[sideblue](\x,\y) circle(2.3pt);}}
  \foreach \x in {2,3}{\foreach \y in {0,1,2,3}{\fill[sideorange](\x,\y) circle(2.3pt);}}
  \foreach \x/\name in {0/a_0,1/a_1,2/a_2,3/a_3}{\node[above=3pt] at (\x,3){$\name$};}
  \foreach \x/\name in {3/b_0,2/b_1,1/b_2,0/b_3}{\node[below=3pt] at (\x,0){$\name$};}
  \node[sideblue] at (.45,1.5){$\overline X_1$};
  \node[sideorange] at (2.5,1.5){$X_1$};
  \node at (1.65,-1.18){(a)};
  \begin{scope}[shift={(7.5,1.55)}]
    \foreach \i in {0,...,7}{\coordinate (v\i) at ({112.5-45*\i}:1.6);}
    \draw[gray] (v0)--(v1)--(v2)--(v3)--(v4)--(v5)--(v6)--(v7)--cycle;
    \draw[cutgreen,line width=1.3pt](v0)--(v1);
    \draw[cutgreen,line width=1.3pt](v4)--(v5);
    \foreach \i in {1,2,3,4}{\fill[sideorange](v\i) circle(2.3pt);}
    \foreach \i in {0,5,6,7}{\fill[sideblue](v\i) circle(2.3pt);}
    \foreach \i in {0,...,7}{\node at ({112.5-45*\i}:1.94){$c_\i$};}
    \node at (0,-2.73){(b)};
  \end{scope}
\end{tikzpicture}
\caption{Cut edges (bisectors) are in green. (a) For consecutive top vertices $a_1,a_2$, $X_1$ consists of the two right columns. It contains a suffix of $A$ and a prefix of $B$. (b) On an eight-cycle, define $X_i$ using consecutive $c_i,c_{i+1}$. Then $X_0=\{c_1,c_2,c_3,c_4\}$ and $X_4$ is its complement. Both cuts are $\{c_0c_1,c_4c_5\}$, showing that an edge can be
counted twice in the sum $\sum_i|\partial(X_i)|$.}
\label{fig:bisectors}
\end{figure}

\section{Labels for two facial subpaths}\label{sec:subpaths}
In this section we prove the following lemma, which is the main contribution of our paper: 
\begin{lemma}\label{prop:subpaths}
The vertices of $A\cup B$ admit labels of $O(m^{1/3})$ words from which every entry $M(r,c)$ can be decoded using the labels of $a_r$ and $b_c$ alone.
\end{lemma}

We recursively partition the distance matrix $M$ into rectangular blocks. By Lemma~\ref{lem:cuts}(i), the row difference $\Delta_i(j):=M(i+1,j)-M(i,j)$ is $-1$ for $j<\pi(i)$ and $+1$ for $j\ge\pi(i)$. We call $\pi(i)$ its \emph{switch position}; a switch at $0$ or $p'$ means that the difference is constant across all columns. For a block
\[
 \beta=[r_0,r_1]\times[c_0,c_1],
\]
call a row difference $i$ \emph{active} if both rows $i,i+1$ and the switch position $\pi(i)$ lie inside the block (see Figure~\ref{fig:switches}). The set of active rows and its size are denoted:
\[
 N(\beta)=\{i:r_0\le i<r_1,\ c_0<\pi(i)\le c_1\},\qquad q_\beta=|N(\beta)|.
\]
\begin{figure}[htb]
\centering
\begin{tikzpicture}[x=.70cm,y=.50cm,every node/.style={font=\small}]
  \node[anchor=east] at (-.25,6.45){$\Delta_i(j)$};
  \foreach \j in {0,...,8}{\node at (\j+.5,6.45){$\j$};}
  \node[anchor=west] at (9.35,6.45){switch position};
  \foreach \x/\lab in {2.5/{c_0=2},6.5/{c_1=6}}{
    \draw[cutgreen] (\x,6.80)--(\x,6.95);
    \node[above,cutgreen,inner sep=2pt] at (\x,6.95){$\lab$};
  }
  \foreach \i/\s in {0/4,1/1,2/3,3/8,4/6,5/5}{
    \pgfmathsetmacro{\yy}{5-\i}
    \foreach \j in {0,...,8}{
      \ifnum\j<\s
        \fill[sideblue!9](\j,\yy) rectangle +(1,1);
        \node[sideblue] at (\j+.5,\yy+.5){$-1$};
      \else
        \fill[sideorange!10](\j,\yy) rectangle +(1,1);
        \node[sideorange] at (\j+.5,\yy+.5){$+1$};
      \fi
      \draw[gray!35,line width=.3pt](\j,\yy) rectangle +(1,1);
    }
    \draw[black!65,line width=1.2pt](\s,\yy+.12)--(\s,\yy+.88);
  }
  \draw[cutgreen,line width=1.3pt](2,1) rectangle (7,5);
  \foreach \i/\s in {2/3,4/6}{
    \pgfmathsetmacro{\yy}{5-\i}
    \draw[cutgreen,line width=2.1pt](\s,\yy+.10)--(\s,\yy+.90);
    \fill[cutgreen](\s,\yy+.5) ;
  }
  \node[anchor=east] at (-.25,5.5){$i=0$};
  \node[anchor=east,cutgreen] at (-.25,4.5){$r_0=1$};
  \node[anchor=east,cutgreen] at (-.25,3.5){$i=2$};
  \node[anchor=east] at (-.25,2.5){$i=3$};
  \node[anchor=east,cutgreen] at (-.25,1.5){$r_1-1=4$};
  \node[anchor=east] at (-.25,.5){$r_1=5$};
  \foreach \i/\s/\reason in {
      0/4/{outside row range},1/1/{left of block},
      3/8/{right of block},5/5/{outside row range}}{
    \node[anchor=west,black!65] at (9.35,5.5-\i){$\pi(\i)=\s$};
    \node[anchor=west,font=\footnotesize,black!65] at (11.6,5.5-\i){\reason};
  }
  \foreach \i/\s in {2/3,4/6}{
    \node[anchor=west,cutgreen] at (9.35,5.5-\i){$\pi(\i)=\s$};
    \node[anchor=west,cutgreen] at (11.6,5.5-\i){active};
  }
  \node[below,cutgreen,inner sep=4pt] at (6.5,-.25){
    $\beta=[1,5]\times[2,6]\qquad N(\beta)=\{2,4\}\qquad q_\beta=2$};
\end{tikzpicture}
\caption{Switches and active row differences. The table shows $\Delta_i(j)=M(i+1,j)-M(i,j)$. The green rectangle contains the differences used by $\beta$: rows $r_0\le i<r_1$ and columns $c_0\le j\le c_1$. Vertical bars mark switches; the two green bars satisfy $c_0<\pi(i)\le c_1$ within this row range. Thus $N(\beta)=\{2,4\}$ and $q_\beta=2$. A switch in the column range is inactive if either matrix row $i$ or $i+1$ lies outside $\beta$.}
\label{fig:switches}
\end{figure}

\begin{figure}[htb]
\centering
\begin{tikzpicture}[x=1.25cm,y=.7cm,every node/.style={font=\small}]
 \fill[cutgreen!12](0,2) rectangle (3,4);
 \fill[cutgreen!12](3,0) rectangle (6,2);
 \draw (0,0) rectangle (6,4);
 \node[align=center] at (1.5,3){opposite sides\\use cut distances};
 \node[align=center] at (4.5,1){opposite sides\\use cut distances};
 \node[align=center] at (4.5,3){same side $0$\\recurse on $\beta_0$};
 \node[align=center] at (1.5,1){same side $1$\\recurse on $\beta_1$};
 \node[above] at (1.5,4){$c<s$};
 \node[above] at (4.5,4){$c\ge s$};
 \node[left] at (0,3){$r\le i^*$};
 \node[left] at (0,1){$r>i^*$};
 \node[below] at (3,0){$s=\pi(i^*)$};
\end{tikzpicture}
\caption{A recursive split. The shaded blocks are answered immediately through the chosen cut. Each row and each column enters exactly one child, and each child has at most $3q_\beta/4$ active row differences.}
\label{fig:split}
\end{figure}
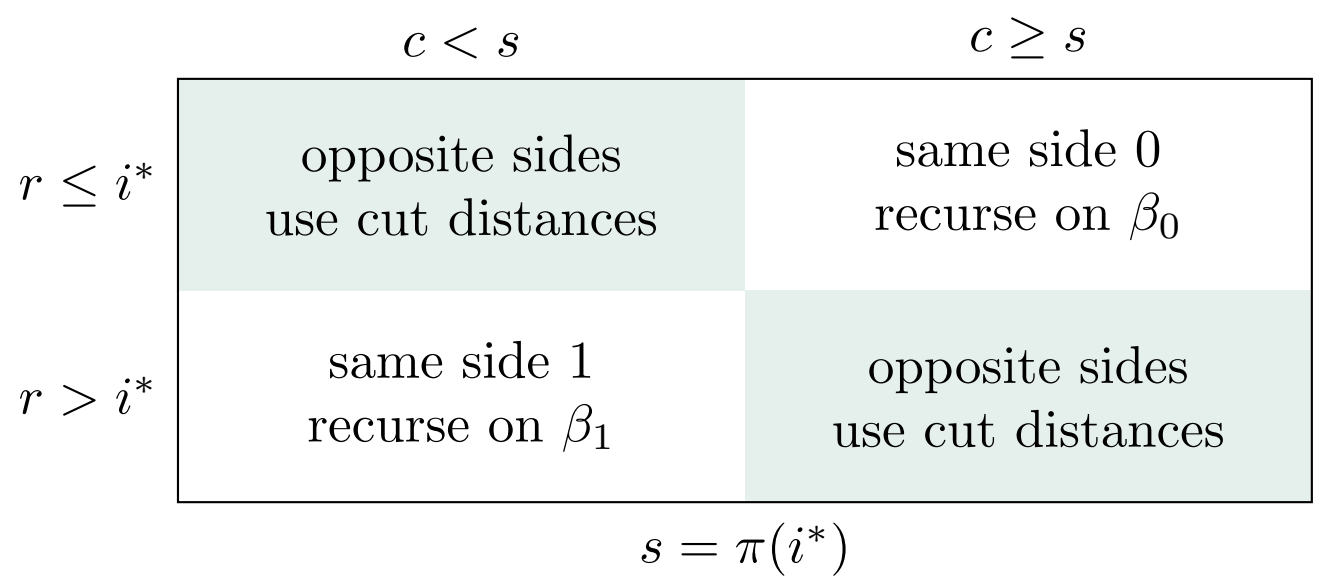

The recursion's root is the full matrix. Set $\tau=m^{2/3}$. A block with $q_\beta\le\tau$ is a leaf. Otherwise, we split the block both horizontally and vertically (into overall four parts) at a chosen active switch $\pi(i^*)$ of some row $i^*$ to be determined later. We answer two of the resulting rectangles by explicitly storing distances to cut-edge endpoints, and recurse on the other two, as detailed below. See Figure \ref{fig:split}. Throughout the recursion, all distances and cuts are taken in the original graph \(G\); only the matrix blocks change.

\subsection{Leaves: count the active switches}
For $(r,c)\in\beta$, summing the row differences in Lemma~\ref{lem:cuts}(i) gives
\begin{equation}\label{eq:count}
{\ M(r,c)=M(r_0,c)+\bigl(M(r,c_0)-M(r_0,c_0)\bigr)
       +2|\{i\in N(\beta):i<r,\ \pi(i)\le c\}|\ }
\end{equation}
Indeed, subtract the sums at columns $c$ and $c_0$. A row contributes two exactly when $c_0<\pi(i)\le c$, and zero otherwise.
This is a dominance-counting labeling problem, as in Abboud, Gawrychowski, Mozes and Weimann~\cite[Section~6]{agmw}. Write $q=q_\beta$ and $w=\max\{1,\lceil\sqrt q\rceil\}$. Order the active indices as $i_0,\ldots,i_{q-1}$ by increasing $(\pi(i),i)$, and partition this list into consecutive groups of $w$ indices.

\begin{itemize}
\item Row $r$ stores its index, $M(r,c_0)-M(r_0,c_0)$, and the prefix counts
\[
 R_r(g)=|\{t<gw:i_t<r\}|,\qquad 0\le g\le\lfloor q/w\rfloor.
\]
\item Column $c$ stores $M(r_0,c)$, $g=\lfloor\rho(c)/w\rfloor$, and the list
\[
 i_{gw},\ldots,i_{\rho(c)-1},\qquad
 \rho(c)=|\{i\in N(\beta):\pi(i)\le c\}|.
\]
\end{itemize}
The required count is $R_r(g)$ plus the number of indices below $r$ in the column's list. Each record uses $O(\sqrt q+1)$ words: the row has $O(q/w+1)$ counts and the column has fewer than $w$ indices. The same formulas handle $q=0$.

\subsection{Large blocks: use a small cut}
Suppose $q=q_\beta>\tau$, and list $N(\beta)=\{n_0<\cdots<n_{q-1}\}$. We wish to choose $i^*$ as the row in which we split the block. We choose the split so that each recursive child contains at most \(3q/4\) active row differences. Thus, we choose $i^*$ among the indices
\[
 J=\{n_t:\lfloor q/4\rfloor\le t\le\lfloor3q/4\rfloor\}.
\]
Specifically, we choose the $i^*$ minimizing $|\cut(X_{i^*})|$. Since $|J|\ge q/2$, Lemma~\ref{lem:cuts}(iii) gives
\begin{equation}\label{eq:smallcut}
 \ell:=|\cut(X_{i^*})|\le\frac{4m}{q} <4m^{1/3}.
\end{equation}

Let $s=\pi(i^*)$ be the column that we split at. Let  $X=X_{i^*}$, and order the cut edges as $x_hy_h$, $1\le h\le\ell$, with $x_h\notin X$ and $y_h\in X$.

Each label stores one bit indicating whether its vertex belongs to $X$.
For a row vertex $a_r$, this bit is $1$ if $r>i^*$ and $0$ otherwise.
For a column vertex $b_c$, it is $1$ if $c<s$ and $0$ otherwise.
By Lemma~\ref{lem:cuts}(i)--(ii), the stored bit is $1$ if the
corresponding vertex belongs to $X$, and $0$ otherwise.
A vertex on side zero stores its distances to $x_1,\ldots,x_\ell$; a vertex on side one stores its distances to $y_1,\ldots,y_\ell$, in the same edge order.
If the two side bits differ, decode by
\begin{equation}\label{eq:cutdecode}
 M(r,c)=\min_{1\le h\le\ell}(\rho_h+1+\gamma_h),
\end{equation}
where $\rho_h$ and $\gamma_h$ are the stored distances. Every candidate is a walk from $a_r$ to $b_c$, so it cannot underestimate the distance. A shortest path crosses the cut, and one of its crossing edges makes the expression exact.

Pairs on the same side belong to one of the two child blocks:
\[
 \beta_0=[r_0,i^*]\times[s,c_1],\qquad
 \beta_1=[i^*+1,r_1]\times[c_0,s-1].
\]
Both are nonempty because $i^*$ is active. Every active index of $\beta_0$ was active in $\beta$ and is smaller than $i^*$; every active index of $\beta_1$ was active in $\beta$ and is larger. If $i^*=n_t$, their counts are at most $t$ and $q-1-t$, respectively, hence at most $3q/4$.

\subsection{Decoding and label length}
The two children partition the rows of their parent, and also partition its columns. Consequently, each row and each column follows a single chain from the root to a leaf. Its label is the sequence of records on that chain, with record types and lengths.

The decoder starts with the two root records. At a cut block it returns~\eqref{eq:cutdecode} if the side bits differ. If they agree, both next records belong to the same child, so it continues there. At a leaf it evaluates~\eqref{eq:count} using the stored counts and list. No information outside the two labels is needed.

To bound a label, let $q_0,\ldots,q_{T-1}$ be the active counts at its cut blocks, followed by a leaf. If there are no cut blocks, only the leaf cost is needed. Otherwise $q_{T-1}>\tau$ and $q_{t+1}\le3q_t/4$, so
\[
 q_t>(4/3)^{T-1-t}\tau.
\]
By~\eqref{eq:smallcut}, the total number of stored cut distances is at most
\[
 \sum_{t=0}^{T-1}\frac{4m}{q_t}
 <\frac{4m}{\tau}\sum_{j\ge0}(3/4)^j
 =\frac{16m}{\tau}=16m^{1/3}.
\]
The leaf uses $O(\sqrt\tau+1)=O(m^{1/3})$ words. There are $O(\log n)$ records, whose overhead is also $O(m^{1/3})$ because $G$ is connected. This proves Lemma~\ref{prop:subpaths}.

\section{Labels for the entire face}\label{sec:complete}
The proof of Theorem~\ref{thm:main} follows easily. 
First apply Lemma~\ref{lem:reduction}. Write the resulting facial cycle as $C=(c_0,\ldots,c_{k-1})$. Recursively halve its index interval $[0,k)$. At each interval of length at least two, apply Lemma~\ref{prop:subpaths} to its left and right halves. These are two vertex-disjoint subpaths of the facial cycle, and every instance uses distances in the same graph $G$.

The label of $c_j$ stores $k$, $j$, and its labels in the at most $\lceil\log_2 k\rceil$ instances containing $j$, ordered by depth and with their lengths. For $j=j'$, return zero. Otherwise find the first split separating $j$ and $j'$; it is determined by $k,j,j'$ alone. Use the two labels for that instance to obtain $\dist(c_j,c_{j'})$. The total size is
\[
 O(m^{1/3}\log k)\text{ words}
   =O(n^{1/3}\log^2 n)\text{ bits}.
\]

\end{document}